\documentclass[prl,twocolumn,showpacs,nofootinbib]{revtex4}
\newcommand{\be}{\begin{equation}}
\newcommand{\ee}{\end{equation}}
\newcommand{\bea}{\begin{eqnarray}}
\newcommand{\eea}{\end{eqnarray}}
\usepackage{graphicx}
\usepackage{epsfig}
\usepackage{bm}
\usepackage{amssymb}

\def\APJ{{\it Ap. J.} }

\def\ASAS{{\it Astron. Astrophys.} }

\def\IJMP{{\it Int. J. Mod. Phys.} }

\def\MNRAS{{\it Mon. Not. R. Ast. Soc.} }

\def\PTP{{\it Progr. Theor. Phys.} }

\begin{document}
\title{Dynamical Mutation of Dark Energy}

\author{L. R. Abramo$^1$, R. C. Batista$^1$, L. Liberato$^2$
and R. Rosenfeld$^2$}

\affiliation{$^1$ Instituto de F\'{\i}sica, Universidade de S\~ao Paulo\\
CP 66318, 05315-970, S\~ao Paulo, Brazil}
\affiliation{$^2$ Instituto de F\'{\i}sica Te\'orica, Universidade Estadual Paulista\\
R. Pamplona 145, 01405-900, S\~ao Paulo, Brazil}


\begin{abstract}
We discuss the intriguing possibility that dark energy may
change its equation of state in situations where large dark
energy fluctuations are present. We show indications
of this dynamical mutation in some generic models of dark
energy.
\end{abstract}

\preprint{IFT-P.xx/2007}
\today
\maketitle

\section{Introduction}
There is ample evidence that supports the
existence of dark matter (DM) and
dark energy (DE) in the universe \cite{Seljak}. 
Due to its charge neutrality and dust-like equation of state (i.e.,
negligible pressure), dark matter starts to cluster
gravitationally very early in the history of the Universe,
and is crucial for the formation of large scale structure.
Dark energy, on the other hand, becomes relevant
only more recently, and is presumed to be a smooth
component with a negative equation of state in order
to fuel the accelerated expansion of the Universe \cite{Reviews}.

At the background level, dark energy (or any other fluid) is completely
determined by its equation of state $w=p_e/\rho_e$, where $p_e$ is the
pressure and $\rho_e$ is the energy density of dark energy. 
Already at
this level dark energy can affect large scale
structures \cite{Linder05} -- see also \cite{LiberatoRosenfeld}
for the effect in different parameterizations of the equation
of state of dark energy.

However, if dark energy is in fact a manifestation of a dynamical
mechanism such as a scalar field, then it will also develop
inhomogeneities due to its gravitational interactions with itself and
with dark matter \cite{Cobleetal}.
In linear perturbation theory,
besides the energy density perturbation
$\delta\rho$, we need two extra degrees of freedom to
characterize cosmological perturbations:
the pressure perturbation $\delta p$ and the
scalar anisotropic stress $\pi$ \cite{Bardeen,KodamaSasaki}.
Alternatively, one can also use the velocity potential
$\theta=\vec\nabla \cdot \vec{v}$ and
the anisotropic stress \cite{Bertschinger}.

The inhomogeneities of dark energy are often quite small,
particularly in the case of ordinary (canonical) scalar 
field models with almost
$\Lambda$-like behaviour -- that is, when $w \simeq -1$ 
\cite{Cobleetal,DuttaMaor,MotaShawSilk}.
In fact, as $w \rightarrow -1$ the perturbations in all dark 
energy models are suppressed in relation to those of dark 
matter. However, if that is not the case
then the dark energy density contrast
$\delta_e \equiv \delta\rho_e/\rho_e$ can be either
small or large, depending
on the pressure perturbations of dark energy \cite{US}.

Here we present further evidence of an intriguing possibility
which was first pointed out in supergravity-motivated scalar field 
models of dark energy \cite{MotaBruck,NunesMota}: that dark energy can
mutate into a fluid with clustering properties similar to those of dark 
matter. We will show that this effect is a generic feature of dark energy, 
and that it has a simple origin: when
pressure perturbations are large, the effective equation of state
inside a collapsed region can be completely different from
the equation of state of its homogeneous component.

\section {Gravitational collapse with dark energy}

In the following, in order to specify the properties of the dark
energy perturbations
we will neglect the anisotropic stress. Moreover, we will
characterize the pressure perturbation in a simplified manner,
using the so-called effective (or non-adiabatic) sound velocity 
\cite{Hu}, defined as $c_{eff}^2 \equiv \delta p_e/ \delta \rho_e$,
which we will assume to be a function of time only.
Notice that this assumption lacks a formal basis in perturbation
theory,
since $\delta p_e$ is a perturbed variable whose time and spatial
dependences can be, and often are, independent of the
variations of $\delta\rho_e$.
Nevertheless, in a particular gauge (the so-called ``rest frame'' 
of the fluid, where $T^i_0 = 0$), the effective sound speed 
coincides with the phase velocity of linear relativistic 
perturbations, $c_X^2$ \cite{Hu,Mukhanov}.

Describing the pressure perturbation as
$\delta p_e = c_{eff}^2 \, \delta\rho_e$
allows us to treat a wide variety
of dark energy models and, crucially, it also allows us to compute
non-linear structure formation using the Spherical Collapse
model (SC) \cite{GunnGott}. Furthermore, in this case
the SC equations (derived in a simplified
relativistic framework) are identical to the equations of
pseudo-Newtonian cosmology \cite{US2}. This means that the
physics of gravitational collapse of structures such as
galaxy clusters is well described within this framework.

Consider then, in the spirit of the SC
model, a spherically symmetric region of
constant dark energy overdensity (the so-called
``top-hat'' density profile.)
Let us call $\rho^c_e = \rho_e + \delta \rho_e$ and 
$p^c_e = p_e + \delta p_e$ the energy density and
the pressure of this region, which are modified with 
respect to the corresponding background quantitites,
$\rho_e$ and $p_e$ by the perturbations $\delta \rho_e$ and 
$\delta p_e$.
The equation of state $w$ is defined as the ratio of
the total pressure to the total energy density, and hence 
it will be different
for the background and the interior of the collapsed region. A simple
calculation shows that the equation of state
inside the collapsed region, $w^c$, is given by:
\begin{equation}
w^c = \frac{p_e + \delta p_e}{\rho_e + \delta\rho_e} =
w + (c_{eff}^2 - w) \frac{\delta_e}{1+ \delta_e} \; .
\label{deltaw}
\end{equation}
For small density contrasts $|\delta_e| \ll 1$, the equation of 
state inside the overdense 
region does not change appreciably. However, 
if $c_{eff}^2 \not= w$, then in the nonlinear
regime, where $\delta \gtrsim 1$ (halos),
there could be a substantial modification in $w^c$ with respect 
to the background equation of state. 
Even in underdense regions (voids), where 
$\delta \approx -1$, there could be large modifications of the
equation of state.
Hence, in principle dark energy could even
effectively mutate into dark matter inside halos and voids.

The above argument is completely general. What remains to be shown is whether 
there are models in which
this dramatic situation is actually realized.
This requires a non-linear analysis of the evolution of pertubations for two
gravitationally coupled fluids, dark energy and dark matter 
(we will neglect radiation and baryons in what follows).
Unfortunately, at present there are no totally rigorous methods
for performing this analysis -- except in the
case of canonical scalar fields, but even then only
approximately \cite{DuttaMaor,MotaShawSilk}.

Here we employ a generalization
of the SC model for the case of a relativistic fluid with pressure.
In the next section we present the
relevant equations and mention under which
conditions they are equivalent to a pseudo-Newtonian approach.
We will then analyse the evolution of the coupled system
of perturbations for a wide variety of dark energy models for
which the pressure perturbations are characterized by
some homogeneous effective sound speed.

\section{Non-linear evolution}

We define $H = \dot{a}/a$ and $h = \dot{r}/r$ as the expansion rates for the 
background ($a$ is the scale factor) and for the
perturbed region ($r$ is the size of the collapsing region),
respectively.
In what follows we work within two assumptions, namely,
that there is no non-gravitational interaction between DE and DM, 
and that the total energy of both DE and DM 
contained in the collapsed region is constant.
The possibility of including DM-DE interactions in the study of 
structure formation was studied in \cite{NunesMota,MotaManera} and
a discussion of possible outflow of energy from the collapsed region can be found in
\cite{MotaBruck,NunesSilvaAghanim}.

Using the continuity equations for the background and for the
perturbed region for a fluid species $j$:
\begin{equation}
\frac{ \dot{\rho}_j}{\rho_j} = - 3 H (1 + w_j) \quad , \quad
\frac{ \dot{\rho}_j^c}{\rho_j^c} = - 3 h (1 + w^c_j),
\end{equation}
we obtain, for the density contrasts:
\begin{equation}
\dot{\delta}_j = - 3 (1+\delta_j) \left[ h (1 +  w^c_j) -  H (1 +  w_j) \right].
\end{equation}
Obviously, for matter we have $w_m=w^c_m=0$.

Using Eq. (\ref{deltaw}) and that the local expansion rate is 
related to the velocity 
field in the perturbed region by $h = H + \theta/3 a$,
we arrive at:
\begin{equation}
  \dot{\delta}_j + 3 H (c_{j \, eff}^{2} - w_j) \delta_j
  + \frac{\theta}{a}  \left[ (1 +  w_j) + (1 + c_{j \, eff}^{2})
  \delta_j  \right]=0 \; .
\end{equation}
The equation that determines the evolution of $\theta$ comes from
the ``acceleration'' in the perturbed region:
\begin{equation}
\dot{\theta} + H \theta + \frac{\theta^2}{3 a} + \frac{3}{2} H^2 a \left
[ \Omega_m \delta_m + (1+3 c_{eff}^2) \Omega_{e} \delta_{e} \right] = 0.
\end{equation}
Notice that there is only one equation for the peculiar velocity,
even in the case of 2 fluids. This is clearly necessary, because 
in the SC model it is a single
spherically symmetric region that detaches from the background,
with a peculiar expansion rate given by $h = H + \theta/3 a$. 
When there is pressure and pressure gradients, 
relativistic corrections almost
surely break this identity between the velocities of the fluids,
which means that the SC model with a top-hat profile
is inconsistent with a fully relativistic calculation.
The assumption that $c_{eff}^2$ is only time dependent
implies that $\delta p$ is also only time dependent -- which
ultimately guarantees the validity of the top hat SC model.

The SC equations capture
many features of the gravitational physics one expects to find at
the scale of the collapsed structures we see today. This is because
the exact same equations can be derived from a pseudo-Newtonian 
treatment of perturbations 
when gradients of pressure can be neglected -- as is the present 
case, of a top hat profile.
When linearized, the ensuing equations correspond to the
linear equations from General Relativity in the case
$c_{eff}^2 = 0$ for sub-horizon scales \cite{Ribamar}.
Furthermore, even
when $c_{eff}^2 \not= 0$, although the equations are not
equivalent anymore, still the growing modes of the 
linearized pseudo-Newtonian perturbations
are identical to the growing modes of the 
linearized relativistic perturbations
\cite{US2}.

\section{Numerical solutions}

We solved the coupled differential equations
for $\delta_{e}$, $\delta_{m}$ and $\theta$, for a few
representative models of dark energy.
In our approximation, a dark energy model is determined by
its background equation of state and the effective speed of
sound.
In a previous paper we investigated the particular case
where the sound speed of dark energy is equal to
its equation of state, $c_{eff}^2 = w$,
and therefore $w^c  = w$. In that case, one can see clearly
from Eq. (\ref{deltaw}) that there is no mutation \cite{US}.

In this letter we expand our previous analysis
to the following cases:
$c_{eff}^2 = 1$,
$c_{eff}^2 = 0$,
$c_{eff}^2 = -1$ and
$c_{eff}^2 = -w$.
The first case is motivated by the common situation when  
dark energy is modelled by a canonical scalar field, 
since in the gauge corresponding 
to the rest frame of the scalar field the effective sound speed is
$c_{eff}^2 = 1$  \cite{Hu}. 
The second case (null pressure perturbations) can occur in
so-called ``silent quartessence'' models \cite{silent}.
The third case represents a perturbation with behaviour close to a cosmological constant 
and the last case reproduces the so-called
generalized Chaplygin gas in a certain limit  ($\alpha = 1$)  \cite{luca}. 

\begin{figure}
\vspace{0.5cm}
\includegraphics[width=9cm]{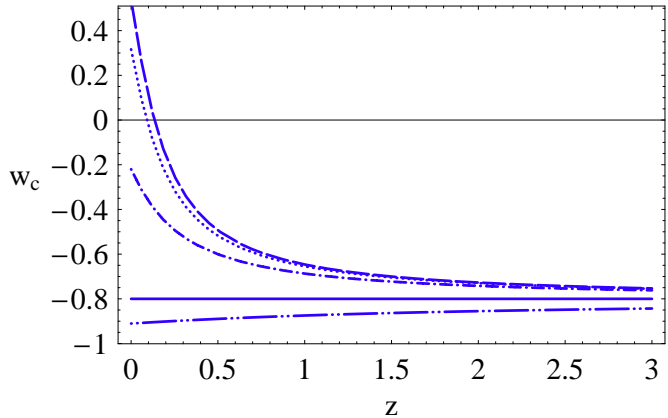}%
\caption{Clustered dark energy equation of state for 
different cases of effective 
speed of sound for background $w = -0.8$:
$c_{eff}^2 = w$ (solid line), $c_{eff}^2 = 1$ (dashed line), 
$c_{eff}^2 = 0$ (dot-dashed line), 
$c_{eff}^2 = -w$ (dotted line) and
$c_{eff}^2 = -1$ (double dot-dashed line).
The instant of turnaround ($h=0$) is $z=0.3$ for $c_{eff}^2 = w$, $z=0.5$ for both
$c_{eff}^2 = -w$ and $c_{eff}^2 = 1$, and $z=0.6$ for $c_{eff}^2 = 0$ (for $c_{eff}^2 = -1$ 
there is no turnaround.)
\label{wc08} }
\end{figure}

\begin{figure}
\vspace{0.5cm}
\includegraphics[width=9cm]{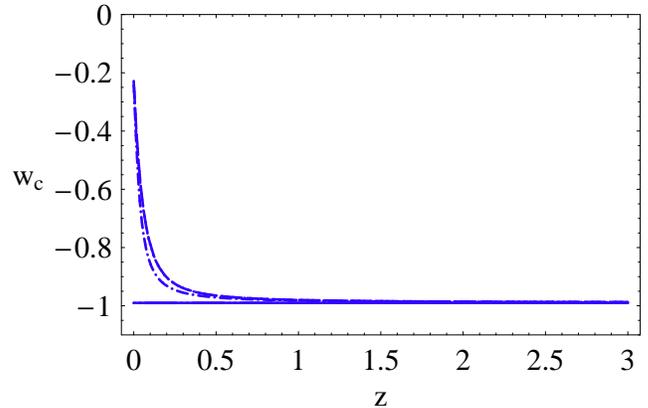}%
\caption{Clustered dark energy equation of state 
for different cases of effective 
speed of sound for background $w = -0.99$. Lines are the same as in Figure 1.
The cases of  $c_{eff}^2 = w$ and $c_{eff}^2 = -1$, as well as 
$c_{eff}^2 = 1$ and $c_{eff}^2 = -w$, 
lie on top of each other.
Turnaround occurs at approximately $z\simeq0.6$ in all cases.
 \label{wc099} }
\end{figure}

We use adiabatic initial conditions for the perturbations
[$\delta_{e}^i = (1 + w) \delta_{m}^i$] at a redshift $z=1000$,
an initial velocity field coincident with the Hubble flow ($\theta^i = 0$),
$H_0 = 72$ km s$^{-1}$ Mpc$^{-1}$, $\Omega_m = 0.25$ and
$\Omega_{e} = 0.75$. In the examples shown below
we fixed the background equation of state of dark energy
at $w = -0.8$.

In Fig. 1 we show the values of the clustered equation
of state $w_c$ as a function of the redshift.
The initial condition $\delta_{m}^i$ was chosen such that the dark energy 
perturbation $\delta_{e} \sim {\cal{O}}(1)$ today.
The perturbation in dark matter is typically one to two 
orders of magnitude larger, which is consistent with the
typical density contrast in galaxy clusters, for which
$\delta_m \sim {\cal{O}}(10^3)$.

As discussed above, for the case $c_{eff}^2 = w$ we obtain no 
mutation in the equation of state in the
perturbed region.
However, substantial modifications are found for other possibilities of 
$c_{eff}^2$. The largest modifications arise for
$c_{eff}^2 = 1$, which in our example is very similar to $c_{eff}^2 = -w$. 
In these two cases, even a complete
metamorphosis of dark energy into a fluid which 
clusters as strongly as dark matter
is possible at recent epochs, due to the large perturbations inside
the collapsed region.
The behaviour for $c_{eff}^2 = -1$ is also easily understood from
Eq. (\ref{deltaw}), since this is the only case where $c_{eff}^2 - w <1$.

As expected, the effect of mutation is greatly reduced for a background 
equation of state close to that of a cosmological constant,
independent of $c_{eff}^2$. 
We illustrate that fact in Fig. 2, where we show the equation of state 
inside the collapsing region
for different models of clustered dark energy, in the case of a background
equation of state $w= -0.99$, and with the same initial conditions 
that were used in Fig. 1. Nevertheless, even in this case
large density contrasts can still arise in the dark energy component,
which lead to the mutation of dark energy.

\section{Conclusions}

We have shown that it is possible to change radically
the clustering properties of dark energy in 
collapsed regions (halos and voids.)
We exemplified this behaviour with a few models for the
dark energy perturbations, and showed that it happens
not only in scalar field models, but also in generic
models of dark energy -- in particular the Generalized
Chaplygin Gas and Silent Quartessence models.

Since the physics of most observed collapsed structures, such as
galaxy clusters, is well approximated by
quasi-Newtonian physics, this dynamical mutation should be
a general phenomenom.
Clearly, this is a crucial issue for all attempts to
compute the influence of dark energy on the formation of 
large scale structures.
More detailed studies, including a relativistic approach and
using different realistic parameterizations of the dark energy
equation of state are currently under way
\cite{US2}.

\section*{Acknowledgments}

We would like to thank
Ioav Waga for many fruitful discussions.
This work has been supported by
FAPESP grants 04/13668-0 (L.R.A. and R.R.) and 05/00554-0 (R.C.B.),
a CNPq grant 309158/2006-0 (R.R.) and a CAPES grant (L.L.).

\end{document}